\documentclass[manuscript]{aastex}

\slugcomment{Submitted to ApJ}

\shorttitle{Deep Near-IR Imaging of $\rho$ Oph Cloud Core}
\shortauthors{Marsh et al.}

\begin{document}

\title{Deep Near-Infrared Imaging of the $\rho$ Oph Cloud Core: \\
    Clues to the Origin of the Lowest-Mass Brown Dwarfs}


\author{Kenneth A. Marsh, Peter Plavchan, J. Davy Kirkpatrick, \\
Patrick J. Lowrance, Roc M. Cutri,}
\affil{Infrared Processing and Analysis Center, California Institute of 
Technology 100-22, Pasadena, CA 91125;
\email{kam@ipac.caltech.edu, plavchan@ipac.caltech.edu, davy@ipac.caltech.edu,
lowrance@ipac.caltech.edu, roc@ipac.caltech.edu}}

\and

\author{Thangasamy Velusamy}
\affil{Jet Propulsion Laboratory, MS 169-506, 4800 Oak Grove Drive, Pasadena,
CA 91109; \email{Thangasamy.Velusamy@jpl.nasa.gov}}

\begin{abstract}
A search for young substellar objects in the $\rho$ Oph cloud core region
has been made with the aid of multiband profile-fitting point-source photometry
of the deep-integration Combined Calibration Scan images of the 2MASS extended 
mission in $J,\,H$ and $K_s$
bands, and {\em Spitzer\/} IRAC images at 3.6, 4.5, 5.8 and 8.0 $\mu$m.
The field of view of the combined observations was $1^\circ\times 9.3'$, and 
the 5 $\sigma$ limiting magnitude at $J$ was 20.5.
Comparison of the observed SEDs with the predictions of the COND and
DUSTY models, for an assumed age of 1 Myr, supports the identification of many
of the sources with brown dwarfs, and enables the estimation of effective
temperature, $T_{\rm eff}$.  
The cluster members are then readily distinguishable from background stars by 
their locations on a plot of flux density versus $T_{\rm eff}$.
The range of estimated $T_{\rm eff}$ values extends down to $\sim750$ K
which, based on the COND model, would suggest the presence of
objects of sub-Jupiter mass.  The results also suggest that
the mass function for the $\rho$ Oph cloud resembles that of the
$\sigma$ Orionis cluster based on a recent study, with both rising
steadily towards lower masses.
The other main result from our study is the apparent presence of
a progressive blueward skew in the distribution
of $J\!-\!H$ and $H\!-\!K_s$ colors, such that the blue end of the range 
becomes increasingly bluer with increasing magnitude.  We suggest that
this behavior might be understood in terms of the `ejected stellar
embryo' hypothesis, whereby some of the lowest-mass
brown dwarfs could escape to locations close to the front edge of the
cloud, and thereby be seen with less extinction.
\end{abstract}


\section{Introduction}

Although it is widely
assumed that brown dwarfs represent the low-mass extension of normal
star formation \citep{boss01}, many of the details are not clear. 
One theoretical problem is to explain the existence of objects much
smaller than the typical Jeans mass of a parent cloud ($\sim1\,M_\odot$),  
especially in light of reports of objects with masses 
$\stackrel{<}{_\sim}3M_J$ \citep{zap02,mar10}. In particular, a possible proto 
brown dwarf of mass $1\!-\!2\,M_J$ has been reported by \citet{bar09}, although
this finding has been refuted by \citet{luh10}.  In principle,
such objects could be produced by
the fragmentation of protostellar cores, which can collapse gravitationally 
provided there is sufficient radiative cooling to overcome
gas pressure.  The lower mass limit for this to occur is believed to
be $\sim1$ $M_J$, irrespective of the details of the fragmentation
process \citep{whit06}.  Disks may play a significant role, either
by gravitational fragmentation of single disks \citep{stam09a,stam09b}, or by
encounters between disks \citep{shen06}.  
One potential difficulty with any of these models is that even if low
mass objects are produced initially, subsequent accretion from the
infalling cloud will increase the final mass
of the object.  This led \citet{reip01} to propose the `ejected stellar
embryo' model whereby the lowest-mass members of a multiple system
are ejected due to dynamical instability before they can accrete
sufficient material to begin hydrogen burning.  Hydrodynamic simulations of
small stellar clusters suggest that dynamical interactions play 
a key role in the formation of not only brown dwarfs, but stars in general
\citep{clarke04}.

Observational estimates of the mass function provide important
constraints on the above models.  Such information can be gained 
using data from deep imaging of
star formation regions at multiple wavelengths. Because brown dwarfs 
emit most of their radiation at wavelengths longward
of 1 $\mu$m, the near infrared is particularly conducive to their 
detection (\citet{lucas00}, \citet{lucas05}).  Such data can be 
compared with the spectral predictions of evolutionary models, 
enabling preliminary estimation of physical 
parameters such as mass 
and temperature.  One model of interest is the dust-free ``COND" 
model of \citet{bar03}, which assumes that the dust grains have settled 
below the photosphere. It thus 
represents one extreme with respect to the absorption of photospheric
radiation by dust, and is applicable to the ``methane dwarfs" whose 
temperatures are $\stackrel{<}{_\sim}1500$ K.  
At the other extreme is the ``DUSTY" model \citep{chab00}, applicable to 
brown dwarfs at higher temperatures, in which the photospheric dust stays 
at its formation site in the atmosphere.

The $\rho$ Oph cloud core is a suitable site for the study of
young brown dwarfs due to its youth, high rate of low-mass star formation 
and its relative proximity; recent estimates of its distance are
$120\pm5$ pc \citep{loin08} and $135\pm8$ pc
\citep{mama08}.  
We assume a distance of 124 pc, representing a
weighted average of these estimates.
The main cloud, L1688, is approximately $1\times2$ pc
in extent and contains a substantial number ($\sim100$) of low-mass
stars \citep{wil05} whose estimated age is $\sim1$ Myr 
\citep{luh99,pra03,wil05}.
Here we describe the results of deep imaging of a dense portion of this cloud 
at multiple wavelengths in the near-infrared and explore the implications for 
brown dwarf formation models.  

\section{Observations and Data Processing}

Our dataset includes the deep $J,\,H$ and $K_s$ images from the $\rho$ Oph
calibration field of the Two Micron All-Sky Survey (2MASS) Extended Mission 
\citep{cutri06} that were produced by combining 1582 observations made
between 1997 and 2000 during the course of the main survey \citep{skrut06}. 
The result is a set of high dynamic range images in which the
sensitivity has been increased by 3.5--4.5 magnitudes over that of the main
survey.  The 2MASS 90009 field, of size $1^\circ \times 9.3'$, centered at 
$\alpha = 16^{\rm h}27^{\rm m}15.6^{\rm s}$, $\delta =-24^\circ41'23''$ 
(J2000), covers part of 
the $\rho$ Oph cloud core, and has limiting magnitudes (at the 5 $\sigma$ level)
of 20.5, 20.0, and 19.0 at $J,\,H$ and $K_s$, respectively. Some additional 
information on the 2MASS Calibration Fields, and on this field in
particular, is given by \citet{plav08a,plav08b}.

We supplement these images with archival data at 3.6, 4.5, 5.8, and 8.0 
$\mu$m from IRAC \citep{faz04} on the {\em Spitzer\/} Space Telescope 
\citep{werner04}, based on observations made in 2004 Apr 27--May 7 as part of 
the c2d legacy project \citep{evans03}.  The archival BCD images were 
reprocessed to correct for artifacts (including saturation, column pulldown, 
muxbleed, ``jailbar" effects, instrumental background and pixel-value outliers) 
and mosaicked using the {\em Spitzer\/} Science Center MOPEX software.  

\begin{figure}
\epsscale{1.0}
\plotone{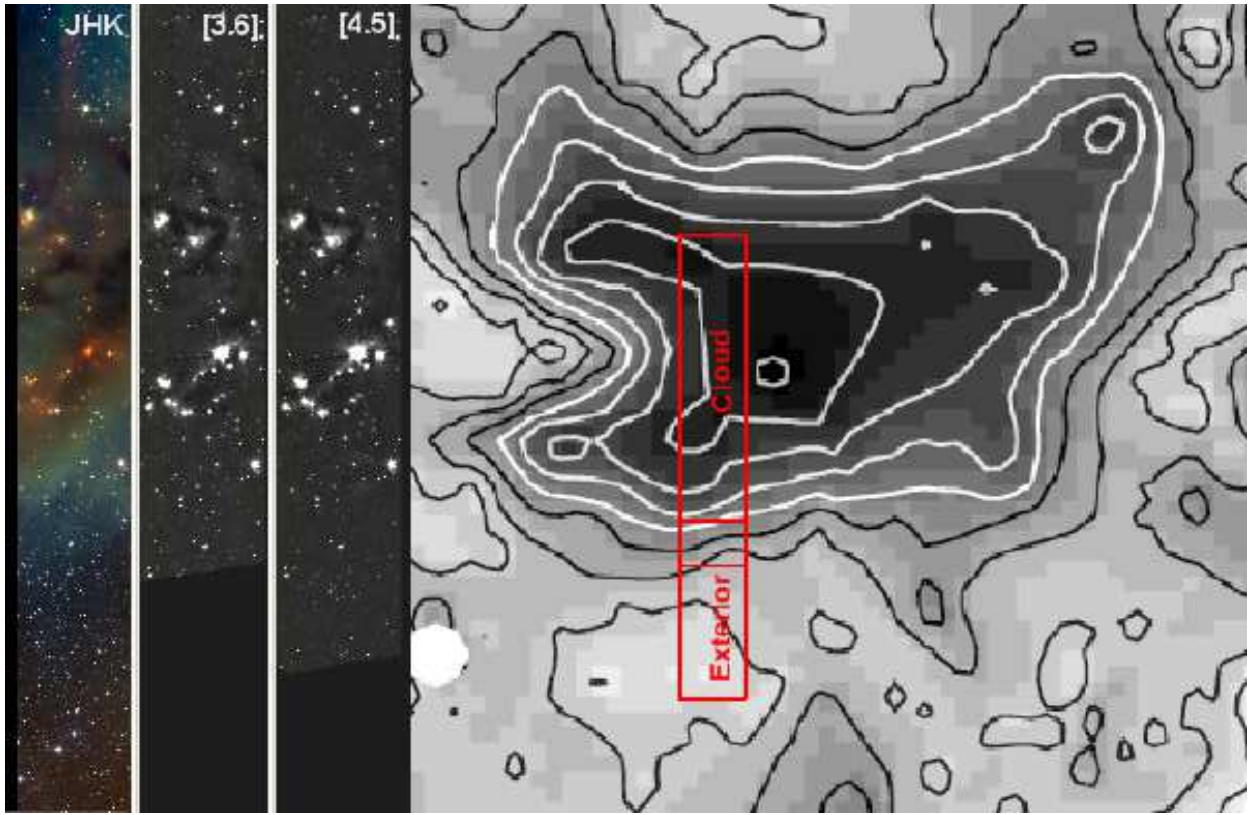}
\caption{The observed region and its relationship to the $\rho$ Oph cloud
core.  The leftmost image is a composite of the 2MASS Deep Field $J,\,H$ and 
$K_s$ images (displayed in blue, green and red, respectively) 
with a field of view of $1^\circ \times 9.3'$, centered on 
$\alpha = 16^{\rm h}27^{\rm m}15.6^{\rm s}$, $\delta =-24^\circ41'23''$ 
(J2000).  To its right are the corresponding 
{\em Spitzer\/} IRAC images at 3.6 and 4.5 $\mu$m, respectively,
obtained by reprocessing {\em Spitzer\/} archival data.
The rightmost section of the figure represents an extinction map from 
\citet{cam99} upon which is superposed an outline of the field being studied
(shown in far-left).  
Also indicated are the two subregions discussed in the text, namely ``Cloud" 
and ``Exterior."}
\label{fig1}
\end{figure}

Figure \ref{fig1} shows the 
relationship of the observed field to an extinction map of the cloud core.
For comparison purposes, the field has been divided into ``cloud" and
``exterior" regions.  The lower boundary of the ``cloud" region was chosen
to coincide with the $A_V=5$ mag contour of
\citet{cam99}, at $\delta=-24.8^\circ$. Our ``exterior"
region was chosen to extend southward from declination $-24.9^\circ$, 
wherein $A_V<3$ mag.  The two regions are therefore defined as follows:
\begin{tabbing}
\quad \= {\em Exterior:\/} \= $16^{\rm h}26^{\rm m}55.0^{\rm s} < \alpha < 
16^{\rm h}27^{\rm m}35.8^{\rm s};\quad -25^\circ11'22'' < \delta < 
-24^\circ54'00''$ \= \kill
 \> {\em Cloud:\/} \>  $16^{\rm h}26^{\rm m}55.0^{\rm s}< \alpha < 
16^{\rm h}27^{\rm m}35.8^{\rm s};\quad -24^\circ48'00''  <  \delta  < 
-24^\circ11'50''$ \\
 \> {\em Exterior:\/}  \> $16^{\rm h}26^{\rm m}55.0^{\rm s}  <  \alpha < 
16^{\rm h}27^{\rm m}35.8^{\rm s};\quad -25^\circ11'22''  <  \delta  < 
-24^\circ54'00''$
\end{tabbing}

The image data from all seven bands were interpolated onto a common pixel-grid 
with a sampling interval of $0.5''$, spatially coincident with the 2MASS 
images.  During this procedure, the {\em Spitzer\/} images were co-registered
with the 2MASS images by matching the positions of bright stars, achieving
a band-to-band registration accuracy of $\sim0.2''$.  These images were
then processed using our MULTIPHOT profile-fitting source 
extraction procedure (Marsh \& Cutri 2010, in preparation), which
represents the multiband extension of the algorithm used for profile-fitting 
photometry in 2MASS.  A key aspect is that the detection step and the
source parameter estimation step are both done using the data at all bands
simultaneously, thereby avoiding the ambiguities involved in bandmerging in
crowded fields.  The parameter estimation step represents a maximum likelihood
solution to the source position ($\alpha,\delta$) and the set of multiband 
fluxes, taking into account the measurement noise and point spread function 
(PSF) at each wavelength.  It provides a set of reduced chi squared values
(for each band individually and an overall value) 
for testing the hypothesis that the
data are consistent with a point source or a blend consisting of a small 
number of closely-spaced point sources.
The noise model itself includes the effects of instrumental noise in
the pixel values, PSF uncertainty, and uncertainty due to confusion error.
The resulting derived error bars for flux and position are believed to 
represent a realistic assessment of the uncertainties in those quantities.

Care was taken to minimize any systematic effects in band-to-band flux
calibration which would compromise the analysis of SEDs, including an
assessment of the effects of confusion.  Based on the
relatively dilute field (on average, about 50 resolution elements
per source) and the accurate band-to-band registration of the images
discussed above, we are confident that the flux variations across bands
(and, in particular, the 2MASS-Spitzer colors) are well-determined.

\section{Results}

\begin{figure}
\epsscale{1.0}
\plotone{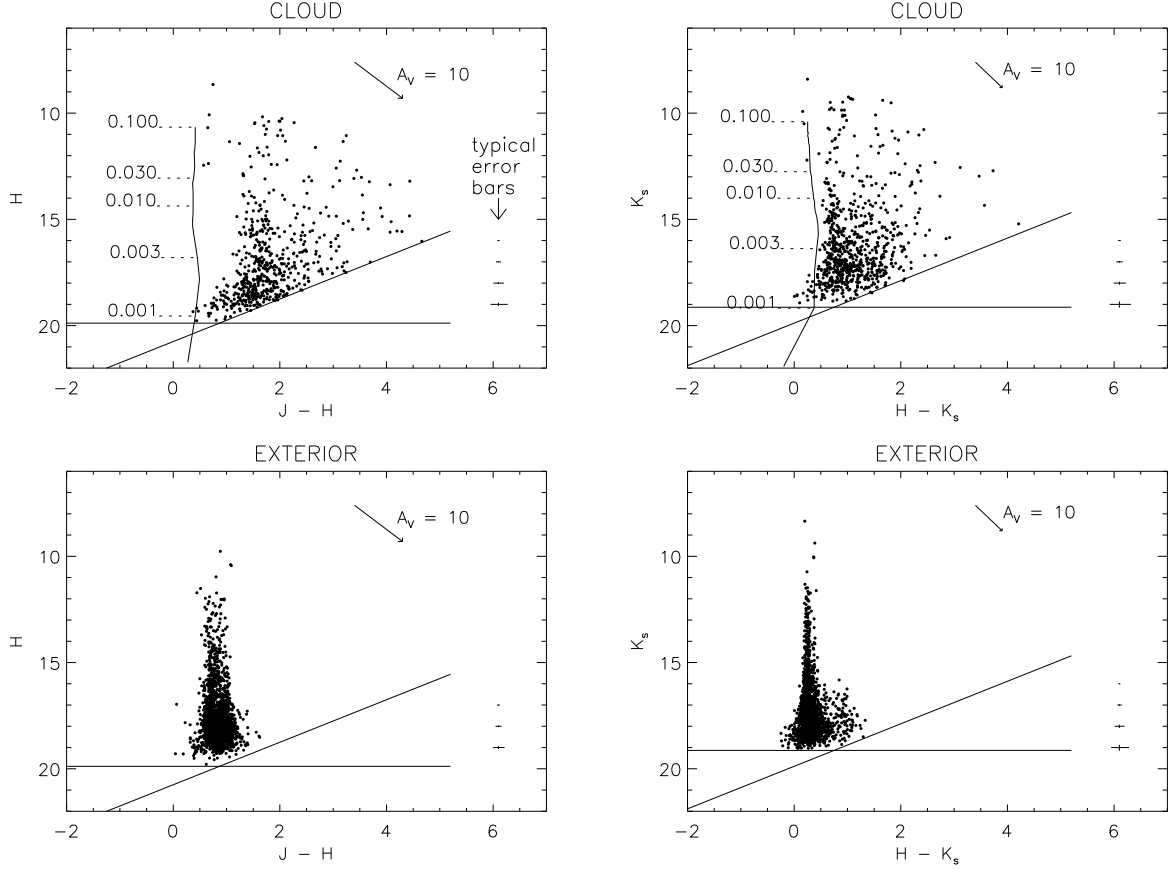}
\caption{Color-magnitude diagrams ({\em left:\/} $H$ versus $J\!-\!H$;  
{\em right:\/} $K_s$ versus $H\!-\!K_s$) for the $\rho$ Oph cloud (upper plots) 
and the region exterior to the cloud (lower plots), where the ``cloud"
and ``exterior" regions are as defined in Figure \ref{fig1}.  The plots
include all detected point sources down to the 5 $\sigma$ level, where the 
sensitivity cutoffs are indicated by the straight solid lines.  
Typical error bars are indicated at the lower right on each plot, each at a
vertical position corresponding to the applicable magnitude.
The arrows represent the reddening vectors corresponding to 10 magnitudes 
of visual extinction.  Also shown for each of the ``cloud" plots is
the 1 Myr isochrone from the COND model \citep{bar03} for the mass values
indicated (in units of solar masses), assuming 124 pc for the cloud distance.}
\label{fig2}
\end{figure}

Figure \ref{fig2} shows color-magnitude diagrams ($H$ versus $J\!-\!H$ and 
$K_s$ versus $H\!-\!K_s$) for the $\rho$ Oph cloud region, and also for the 
region exterior to the cloud for comparison (see Figure \ref{fig1}).  
For both band/color combinations, the ``cloud" 
and ``exterior" plots are clearly quite different; the
plots for the exterior region are consistent with relatively unobscured 
background stars, whereas the plots for the cloud region
are consistent with stars at a wide range of visual obscurations.  

For the $H$ versus $J\!-\!H$ plot of the cloud region, it is also evident that 
for most of the $H$ apparent magnitude range, the leftmost edge of the set of 
$J\!-\!H$ values is separated from the 1 Myr isochrone of the COND model by 
about 10 magnitudes of visual extinction, consistent with the star-forming 
cluster being embedded in the cloud at a depth corresponding to this opacity.
However, for $H\stackrel{>}{_\sim}18$ mag, this ``minimum extinction"
edge turns blueward, reaching zero extinction at $H\simeq 19.2$ mag.
A similar blueward skew is apparent for the corresponding $K_s$ versus 
$H\!-\!K_s$ plot.  We believe this effect is real since it
is substantially larger than the observational error bars and is 
not evident for sources exterior to the cloud, as is apparent from
Figure \ref{fig2}. It is also not an artifact of the profile-fitting source 
extraction process because the same blueward skew is evident when the 
extraction is repeated using aperture photometry via DAOPHOT \citep{stet87}.  
The aperture photometry results are not plotted here, but can be accessed 
via \citet{cutri06}.

If the extracted sources were all members of the star-forming cluster in
the $\rho$ Oph cloud at 124 pc,
then the faintest sources would correspond to low-mass brown dwarfs according
to the COND model. It is likely, however, that many of the
sources are external to the cloud and represent other types of
object along the line of sight.  Possibilities for the latter are more distant 
background stars, foreground T dwarfs,
and extragalactic objects such as AGN.  The first of those possibilities is 
suggested by previous measurements of the $K$-band luminosity function of the
cloud core region, which is dominated by background stars for $K>15$ 
\citep{luh99}.  Such objects would be most visible near the less-extincted
edges of the cloud.  Could they then account for the blueward skew
seen in the color-magnitude diagrams of Figure \ref{fig2}?  The answer
is no, because any low-extinction window in the field would admit
background stars over a wide range of magnitudes, and no blueward-turning 
edge would result.

In order to separate brown dwarfs from background stars we exploit the
fact that these two classes of object are characterized by different color 
temperatures. Accordingly, we have fitted a temperature to each of the 
extracted sources from our multiband photometry.  The technique, described 
in the next section, does not presume cluster membership {\em a priori\/}.

\section{Temperature Estimation}

We have estimated the effective temperature, $T_{\rm eff}$, of each extracted 
source based on a least squares fit of model spectra to the observed SEDs. 
The following suite of models was considered:
\begin{itemize}
 \item [1.] COND \citep{bar03} for an assumed age of 1 Myr.
 \item [2.] DUSTY \citep{chab00} for the same assumed age.
 \item [3.] NextGen \citep{haus99} for stars of solar gravity and solar 
metallicity.
\end{itemize}
We have restricted these fits to five of our seven wavebands, namely 
$J,\,H,\,K_s,\,[3.6]$ and $[4.5]$ because of their close wavelength 
correspondence with published COND and DUSTY model results at the 
$J,\,H,\,K,\,L'$ and $M$ bands.  This restriction does not negatively impact the
temperature estimates since the sensitivity of IRAC at 5.8 and 8 $\mu$m is
significantly less than for the shorter-wavelength bands. In addition,
the photospheric fluxes at 5.8 and 8 $\mu$m are, in some cases, 
significantly contaminated by dust emission from circumstellar
disks \citep{sch07}.

For each hypothesized model, the procedure consisted of minimizing a 
function $\phi(T_{\rm eff},\alpha,A_V)$ with
respect to $T_{\rm eff}$, a flux scaling factor, $\alpha$, and the visual 
absorption, $A_V$, based on a functional form given by:
\begin{equation}
\phi(T_{\rm eff},\alpha,A_V) = \sum_{\lambda=1}^{5} \frac{1}{\sigma_\lambda^2}
[f_\lambda^{\rm obs} - \alpha
10^{-0.4r_\lambda A_V}f_\lambda^{\rm mod}(T_{\rm eff})]^2 - \beta A_V
\label{eq1}
\end{equation}
where $f_\lambda^{\rm obs}$ and $f_\lambda^{\rm mod}(T_{\rm eff})$ represent the
observed and model fluxes, respectively, at the wavelength band represented 
by index $\lambda$; $\sigma_\lambda$ represents flux uncertainty; 
$r_\lambda$ represents the absorption 
in band $\lambda$ relative to $A_V$, based on the reddening law from
\citet{allers06}.   The quantity $\beta$ is a constant whose significance
we now discuss.

The 2nd term on the right hand side of (\ref{eq1}) is designed to
compensate for spurious biases towards low $T_{\rm eff}$ values resulting from the
potential presence of long-wavelength excesses due to circumstellar dust.
It is necessitated by the increasing degeneracy which exists between 
$T_{\rm eff}$ and $A_V$ as the temperature increases and the Rayleigh-Jeans
regime is approached.  Because of this degeneracy, the SED
of a high-temperature object with large extinction can be reproduced 
equally well by a model based on a low-temperature object seen though 
small extinction.
For such an object, the presence of even a relatively small long-wavelength
excess due to circumstellar dust can favor
a spurious low-temperature solution.  To compensate for this effect, we
have imposed a penalty against low values of $A_V$, whose severity is
controlled by the value of $\beta$.  We have optimized the value of
$\beta$ using published photometry for a large number of 
spectroscopically-confirmed brown dwarfs of a variety of ages, as
discussed in Appendix 1; the optimal value was found to be 0.7.  
This optimization incorporated the results of our own spectroscopic
observations of 7 of the objects in our $\rho$ Oph source extraction list
\citep{mar10}.

In evaluating $\phi$, the
model fluxes were calculated by logarithmic 
interpolation of the tabulated model fluxes with respect to $T_{\rm eff}$ and 
wavelength.  The wavelength interpolation was necessitated by the 
mismatch between the observational wavebands (the {\em Spitzer\/} 3.6 $\mu$m
and 4.5 $\mu$m bands) and the $L'$ and $M$ bands for which the model 
photometry was tabulated.  Although the wavelength mismatches are relatively
small, some errors in the interpolation are inevitable.
To assess this, we took some individual high resolution spectra
from the COND models and smoothed them with the 2MASS and {\em Spitzer\/}
bandpasses and found that $K_s\!-\![4.5]$ was accurately reproduced
(to within 0.1 mag or better) but that $K_s\!-\![3.6]$ was less so, particularly
at the lowest temperatures.  Nevertheless, given the other errors involved
in the temperature estimation procedure, we have found that this 
interpolation error is a relatively minor contributor.

\begin{figure}
\epsscale{0.75}
\plotone{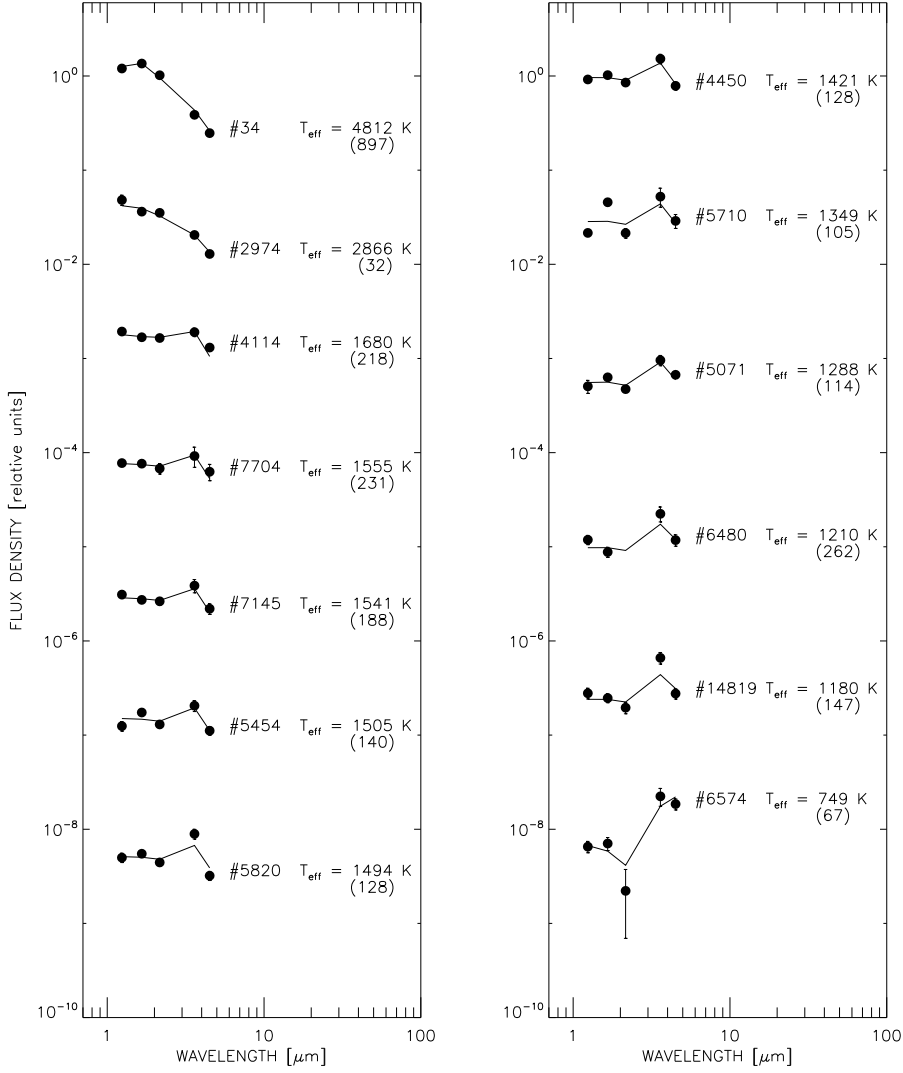}
\caption{Sample spectral fits, representing the results for two of the
hotter objects (\#34 and \#2974) plus the 11 objects with the lowest estimated 
$T_{\rm eff}$. The filled circles with 
error bars represent the dereddened fluxes estimated from the 
observations, and the solid lines represent the best-fit model spectra. 
The latter were derived from the NextGen model
for object \#34, the DUSTY model for object \#2974, and the COND model for the 
remaining 11 objects.  Quantities in parentheses represent the formal
uncertainties of maximum likelihood estimation of $T_{\rm eff}$ [K]. Note,
however, that they
do not take into account the effects of model error (see Appendix).  The
estimated $A_V$ values were (in order of decreasing $T_{\rm eff}$): 10.1, 20.2, 8.4, 1.5, 6.9, 9.2, 9.9, 7.2, 8.5,
11.3, 6.0, 7.3, and 1.1 mag.  The corresponding values of reduced chi squared
were: 1.1, 1.5, 1.3, 1.5, 1.4, 1.1, 1.2, 0.9, 19.7, 1.7, 2.4, 2.9, and 2.1.}
\label{fig3}
\end{figure}

For a given model,
$T_{\rm eff}$ was restricted to the nominal range of physical validity,
corresponding to $T_{\rm eff}<2000$ K (COND),
$1800<T_{\rm eff}<3000$ (DUSTY), or $T_{\rm eff}>2200$ K (NextGen).  
For each extracted source, the estimates of $T_{\rm eff}$, $\alpha$, and $A_V$ were 
then based on the model which gave the minimum value of 
$\phi$.  Such estimates were made for all sources detected in at least 4 of
the 5 bands.  Solutions which gave a poor fit, as indicated by a large 
value of reduced chi squared ($\chi_\nu^2>20$), were excluded.  A total of 1022
successful fits was thereby obtained for the ``cloud" region, representing 
97\% of the extracted
sources.  In particular it was found that the predicted spectral peculiarities
of low temperature brown dwarfs ($T_{\rm eff}\sim1000$ K) were well fit by the 
observations.  Figure \ref{fig3} shows some sample model fits to the 
dereddened fluxes, with estimated $T_{\rm eff}$ values as indicated. 

The temperature fitting procedure was repeated for the ``exterior" 
region.  This region was not covered by the 3.6 $\mu$m observations and
was incomplete at 4.5 $\mu$m, as can be seen from Figure \ref{fig1}. 
Nevertheless, temperature fitting was still possible with relatively little loss
of accuracy.
A total of 684 successful fits was thereby obtained, representing 100\% of the 
extracted sources.
This is comparable to the number of fits obtained for the ``cloud" region even 
though the area of sky was substantially smaller.  We verified that
the lack of 3.5 $\mu$m data did not cause any systematic effects by 
repeating the ``cloud" analysis without the 3.5 $\mu$m data.

\begin{figure}
\epsscale{0.8}
\plotone{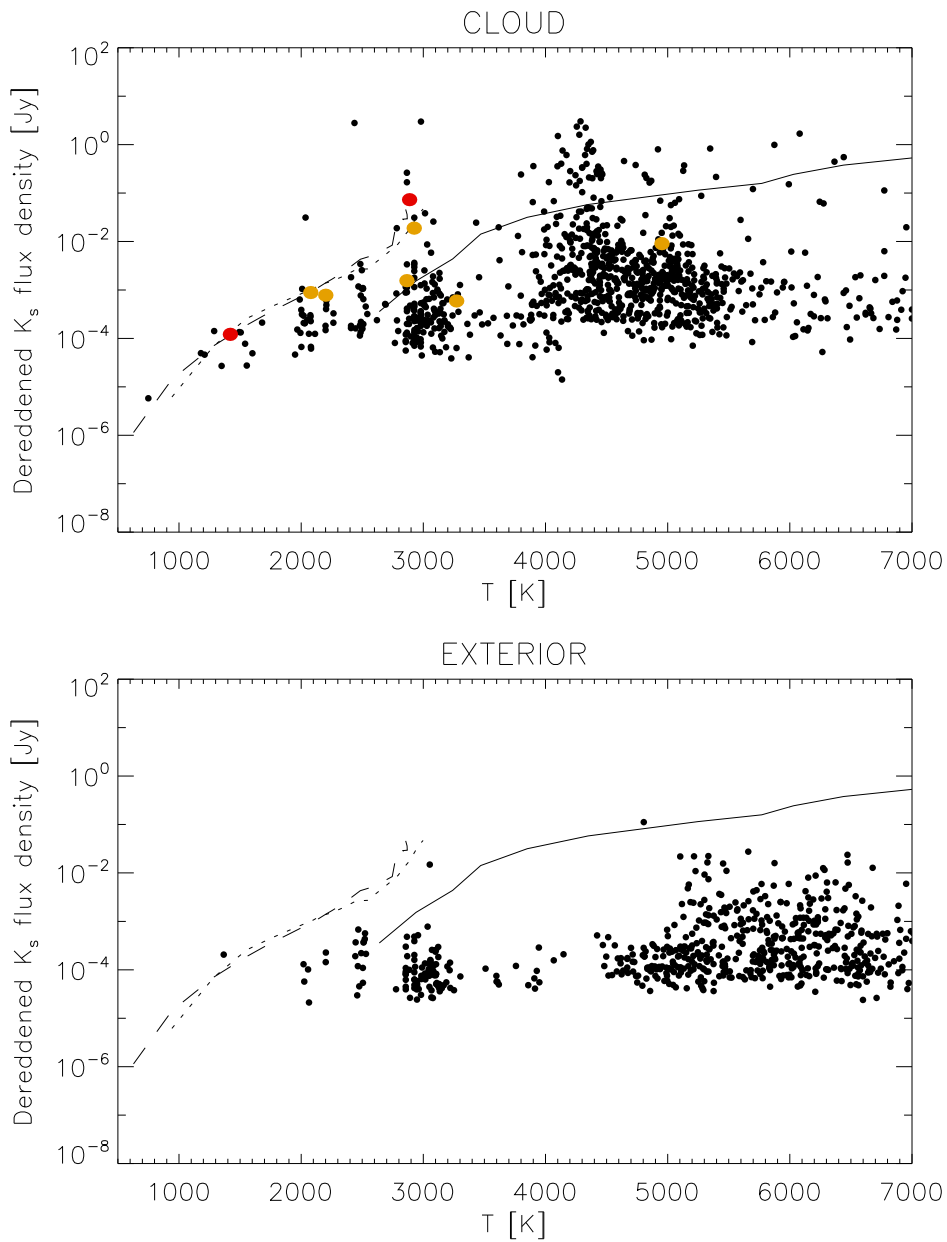}
\caption{Dereddened $K_s$-band flux as a function of estimated temperature,
for the $\rho$ Oph cloud (upper) and the cloud-exterior region (lower).
The red symbols represent the results for the two 
spectroscopically-confirmed brown dwarfs in this region (objects \#60 and 
\#4450 in Table 1), the yellow symbols represent the
six objects observed by \citet{mar10} in addition to \#4450, and the black 
symbols represent all other sources.  Also shown are the 
model curves for the 1 Myr COND (dashed) and DUSTY (dotted) models, and 
main sequence stars (solid) for an assumed distance of 124 pc.}
\label{fig4}
\end{figure}

Figure \ref{fig4} shows plots of dereddened $K_s$ flux as a function of 
estimated temperature for the ``cloud" and ``exterior" regions. Also shown 
are the model curves for the COND and DUSTY models (age 1 Myr) and main 
sequence stars for an assumed distance of 124 pc.  

\section{Cluster membership}

Comparison of the two plots in Figure \ref{fig4} shows that
the number of points which fall on or close
to the COND/DUSTY model curves is much greater on the ``cloud" plot than
on the ``exterior" plot, consistent with the presence of brown dwarfs in
the cloud core region.
It is also apparent that the majority of points in the ``exterior" plot
correspond to $T_{\rm eff}\stackrel{>}{_\sim}2800$ K and are below the main 
sequence line for a distance of 124 pc.
They are therefore consistent with normal stars at distances larger than
124 pc.  The same population is evident in the ``cloud" plot, and hence
the points which fall below the main sequence line on that plot can
confidently be identified as background stars.
They number 882 and 666 in the ``cloud" and ``exterior" regions,
respectively.
After exclusion of those objects we are left with $N_{\rm cloud}=139$ 
cluster member candidates 
in the ``cloud" region and $N_{\rm ext}=18$ in the ``exterior" region.  
This set of candidates may still, however, contain non-cluster contaminants.
Possibilities include:

\begin{itemize}
 \item [1.] Foreground L and T dwarfs.
 \item [2.] Extragalactic objects such as AGN.
 \item [3.] Main sequence background stars whose temperatures have been 
underestimated and which have therefore been incorrectly assigned to the
brown dwarf regime in Figure \ref{fig4}.
\end{itemize}

The likelihood of foreground contamination by L and T dwarfs
can be assessed from available number density statistics given that the 
volume of space in the cone capped by our ``cloud" region is 54 pc$^3$.  
Since the estimated space densities of L and T dwarfs in the solar neighborhood
are $\sim0.01$ pc$^{-3}$ in both cases (\citet{bur07}, \citet{met08}),
the corresponding expectation numbers within this region are $\sim0.5$
for both object types. This is consistent with estimates of the $J<21$ 
magnitude-limited sky density of T dwarfs \citep{bur04} which would suggest
$\sim0.5$-1.0 such objects in a region of this size.  We therefore conclude that
the number of low-$T_{\rm eff}$ foreground objects in the ``cloud" region is
relatively small.

We can obtain an upper limit to the number of extragalactic 
contaminants by assuming that all of the inferred cluster members in the 
``exterior" region are spurious. 
We can then predict the number of contaminating sources 
in the ``cloud" region by scaling $N_{\rm ext}$ by the cloud:exterior 
background source
count ratio\footnote{It is not appropriate to scale by
the relative areas of the two regions, since background source counts 
are heavily influenced by absorption, which is different for the two regions.
The scaling must instead be based on the number density ratio of 
extragalactic sources to background stars, which we
can safely assume is the same for the ``cloud" and ``exterior" regions.},
equal to 1.32 from the background source counts estimated above.
On this basis, an upper limit for the number of non-cluster members included in
$N_{\rm cloud}$ is 24, i.e. of our 139 cluster member candidates, the 
percentage of contaminating sources is between 0 and 17\%.

A particularly interesting subset of the cluster member candidates in 
Figure \ref{fig4} is the group of low-$T_{\rm eff}$ objects ($T_{\rm eff}<1800$ K), of which there
are 11 in the ``cloud" region---these are most likely low-mass brown dwarfs.
By contrast, there is only 1 such object in the ``exterior" region.
The fact that the 
number of low-$T_{\rm eff}$ objects decreases so sharply when going from 
``cloud" to ``exterior" supports their inferred cluster membership and
argues strongly against their being extragalactic. Even the 1 low-$T_{\rm eff}$ 
object in the ``exterior" region could be a foreground T dwarf based
on the statistics quoted earlier.  We therefore consider it unlikely that any
of our low-mass brown dwarf candidates are extragalactic background
sources.  Presumably, the latter are shielded from the observed field by 
dust in the Galactic disk.  

Is it possible that we have underestimated the background contamination
due to the fact that background sources in the ``cloud"
region are much more heavily reddened than those in the ``exterior" region?
For example, do highly reddened AGN resemble brown dwarfs when
viewed through the cloud? To test this, we have simulated the effect of
viewing the ``exterior" region through a dense dust cloud and have repeated our
temperature fits using the artificially reddened data.  We found that no value
of applied $A_V$ could produce a significant population of brown dwarf
false-positives,  and therefore conclude that
the majority of our brown dwarf candidates in the ``cloud" region cannot be 
attributed to reddened background objects.

\begin{figure}
\epsscale{0.45}
\plotone{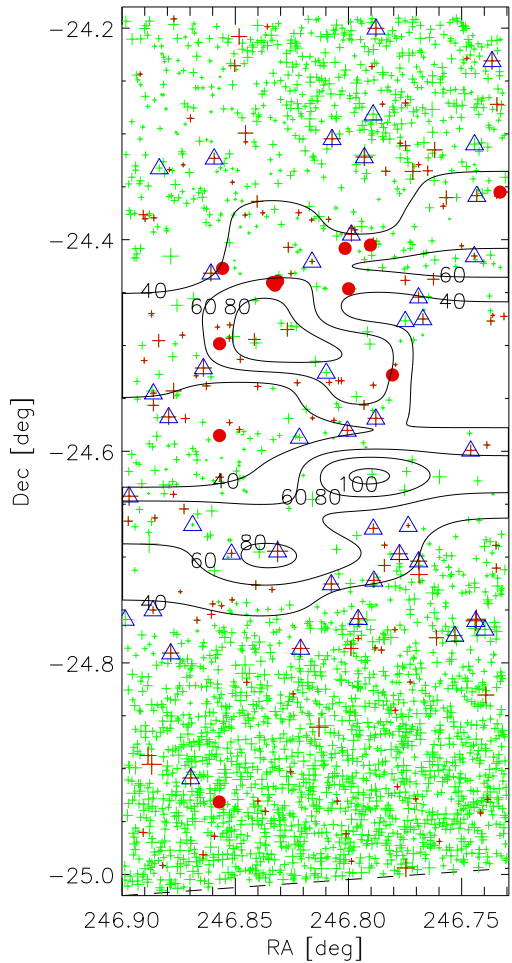}
\caption{The spatial locations of detected sources in the 
$\rho$ Oph cloud. Red symbols denote candidate cluster members and green
symbols denote inferred background objects.  Filled circles represent
candidate low-mass ($<5M_J$) brown dwarfs, and all other objects are
indicated by crosses whose size increases with $K_s$ flux.
Objects which coincide with known T Tauri
stars/YSOs are enclosed by blue triangles. Also shown are contours of $A_V$, 
at levels of 40, 60, 80 \& 100 mag.  These go to deeper absorption levels
than was possible with the optical star count technique used to generate the
extinction contours shown in Figure 1, but are nevertheless consistent with
the contours in that figure.
The dashed line represents the limit of coverage at the IRAC bands.
Note that the aspect ratio of this figure differs from that of 
Figure \ref{fig1} in that the RA axis has been stretched to reduce
the crowding of plotted symbols.}
\label{fig5}
\end{figure}

Having excluded the various classes of possible background objects on the 
above basis, we are left with a total of 165 candidate cluster members in the 
entire observed field, of which 92 are brown dwarf candidates.  These
165 objects represent the sum of the number of objects in the cloud
region (139), exterior region (18), and the gap between those two
regions (8).
The photometric results for the candidate cluster members are presented in 
Table \ref{tbl-1}, and the locations of all detected sources (observed in
at least 4 bands) are plotted in
Figure \ref{fig5}.  In the latter, red and green symbols denote candidate
cluster members and background stars, respectively.  
Also shown for comparison on the plot are our estimated contours of visual
absorption, $A_V$;  these were obtained by dividing the field of view into
square cells of size $2'\times2'$, finding the maximum value of 
$A_V$ for all stars falling into a given cell, and interpolating using a
Gaussian convolution kernel.  

The fact that the background source density in Figure \ref{fig5} is highest 
(by far) in the off-cloud region provides validation of our classification
criterion for background stars. Likewise, the fact that the low-mass 
($<5M_J$) brown dwarf candidates (filled red circles)
are concentrated mostly in the cloud portion provides some confirmation of
their identity as cluster members.  

\begin{figure}
\epsscale{1.0}
\plotone{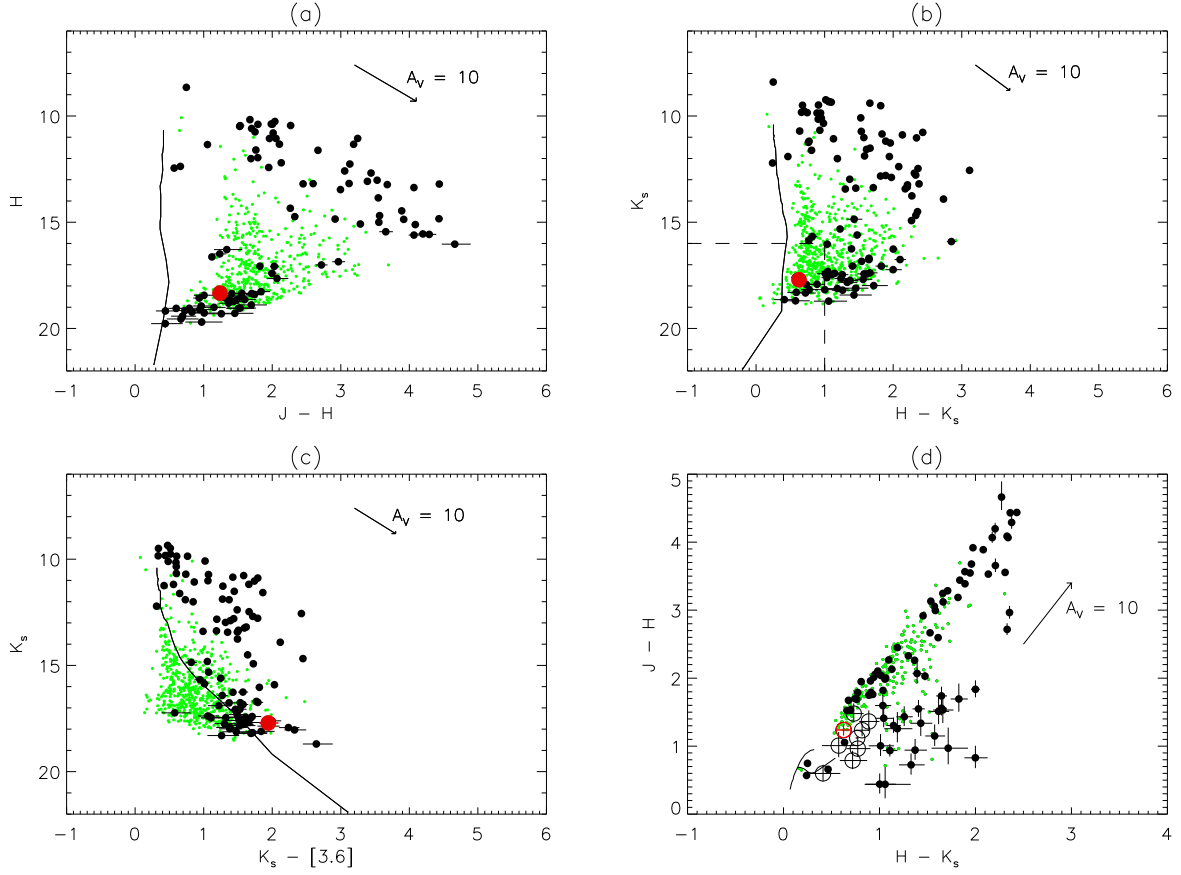}
\caption{Color-magnitude and color-color diagrams for objects 
in the $\rho$ Oph cloud, subject to $S/N\ge5$ at all bands on a given plot.
Red symbols represent object \#4450 (the spectroscopically confirmed
low-mass brown dwarf), and black symbols represent all other cluster members.  
Green symbols represent inferred background stars.
The open circles in the color-color diagram in (d) represent the cluster 
members within the zone delineated by the dashed lines in (b).
Also shown for comparison are 
the predictions of the COND model for age 1 Myr (solid curves in (a)-(c)), and 
the loci of main sequence and giant stars (solid curves in (d)), from 
\citet{bess88}.}
\label{fig6}
\end{figure}

Color-magnitude plots for the candidate cluster members are shown in panels
(a)-(c) of Figure \ref{fig6}.  Also shown, for comparison, in the 
$K_s$ versus $K_s\!-\![3.6]$ plot of panel (c) are the locations of the 
inferred background stars.  Since the $K_s\!-\![3.6]$ color is quite 
temperature-sensitive in the brown dwarf regime, the distributions of 
brown dwarfs and background stars are distinctly different on this plot.  
The fact that the blueward skew is 
still evident in the $JHK_s$ plots of (a) and (b) indicates that it is 
associated with objects in the 
$\rho$ Oph cloud, most likely brown dwarfs.  The well-defined portion of
the blueward skew has been delineated by the dashed lines on the plot;
the objects which fall within this zone have been plotted as open circles in
the $J\!-\!H$ versus $H\!-\!K_s$ color-color diagram of panel (d).
The fact that the colors are positively 
correlated on this plot (i.e., blue $J\!-\!H$ corresponds to blue $H\!-\!K_s$)
confirms that the blueward 
skew is not a random effect caused by increasing measurement errors.
Does the blueward skew then indicate a 
deficiency in the models?  This is unlikely, since the COND model
is based on a dust-free photosphere and thus already produces relatively blue 
colors.  A more likely explanation is that the faintest sources
are seen with the least extinction.  This could occur if
the lowest-mass brown dwarfs are preferentially ejected from
their formation sites and some are therefore seen closer to the front edge
of the cloud along the line of sight.  Such behavior would, in fact, be
expected on the basis of the `ejected stellar embryo' hypothesis, although
models which involve ejection subsequent to the accretion phase would
not be excluded.

An additional feature of the color-magnitude diagrams in Figure \ref{fig6}
is the dip in density of points around $H\sim16$.  Similar drops at 
comparable magnitudes have been found for other young clusters and
star-forming regions \citep{muench03,lucas05,cab07}.  As discussed by
\citet{cab07}, this phenomenon is yet to be understood,
but it may be related to the deficit in M7 and M8 objects
noted by \citet{dob02}.

\section{The Mass Function}

\begin{figure}
\epsscale{1.0}
\plotone{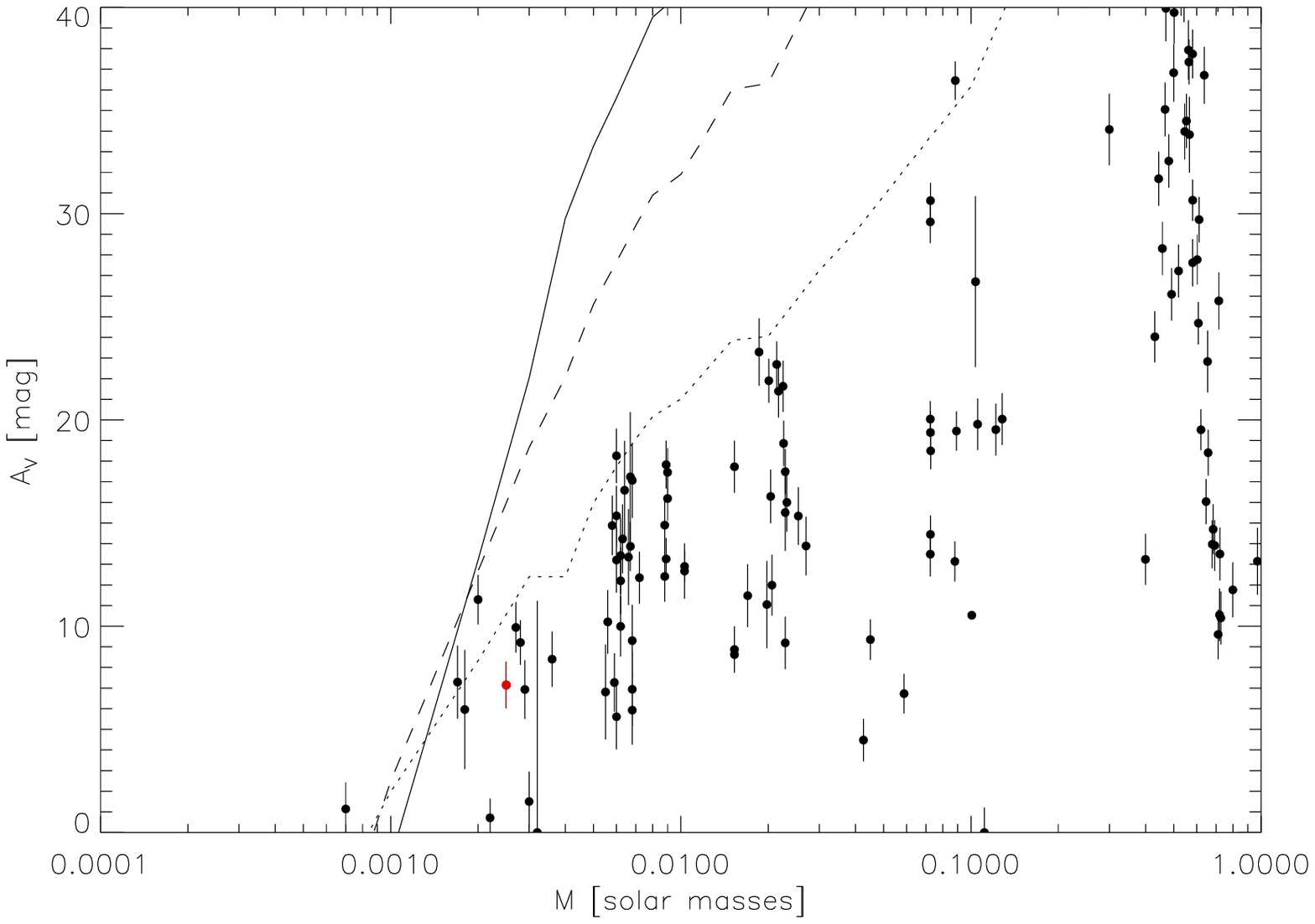}
\caption{Estimated visual extinction as a function of estimated mass for
objects in the $\rho$ Oph cloud. For reference, object \#4450 is indicated in 
red. The 5 $\sigma$ sensitivity cutoffs for 
$J,\,H$ and $K_s$ are indicated by the dotted, dashed and solid lines, 
respectively.}
\label{fig7}
\end{figure}

The temperature estimates can be used to infer a mass for each object in
the ``cloud" region, based on the assumed age.  For those objects which were 
fit by the COND or DUSTY model, each temperature then corresponds to a unique 
model-based mass.
For the hotter objects, which were best fit by NextGen, we used a 
mass-temperature relationship derived from observations by \citet{gre95} 
of pre-main sequence objects in the $\rho$ Oph cloud.  
Figure \ref{fig7} shows a plot of $A_V$ versus mass, $M$, from which it
can be seen that for $M\stackrel{<}{_\sim}0.02\,M_\odot$, the minimum 
extinction for a given 
mass is positively correlated with the mass, reaching zero extinction for 
$M\sim0.003\,M_\odot$ (3 Jupiter masses).  No conclusion can be drawn
for masses below about 0.001 $M_\odot$, however, since sources with
significant extinction would be below the $J,H,K_s$ sensitivity limits for 
that mass range.

\begin{figure}
\epsscale{1.0}
\plotone{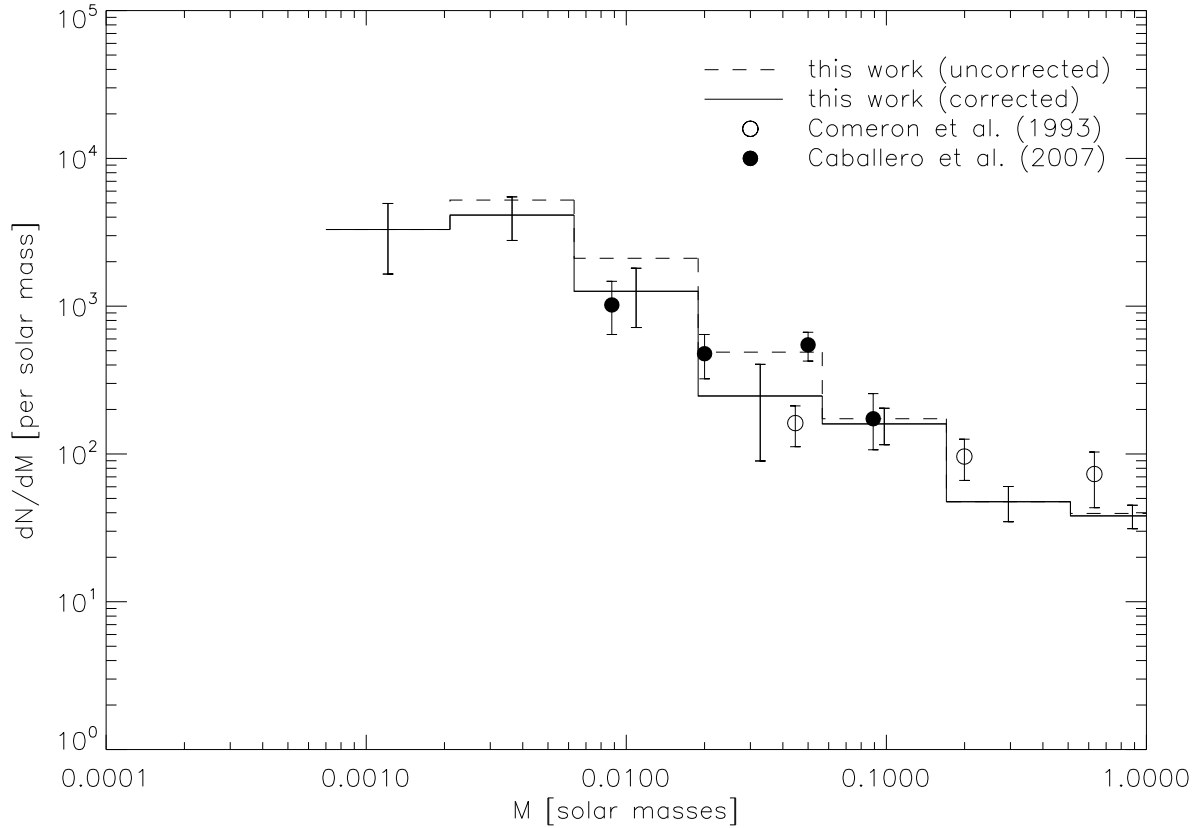}
\caption{The mass function for objects in the $\rho$ Oph cloud. The
solid line (with error bars) represents our best estimate, based on
an assumed age of 1 Myr in conjunction with the COND and DUSTY models.
It has been corrected for 
the expected effects of contamination by foreground and background objects.
The dashed line represents the estimate prior to applying the correction.
Also shown for 
comparison are the estimated mass functions of $\rho$ Oph \citep{com93}
(open circles) and $\sigma$ Ori \citep{cab07} (filled circles) scaled 
on the vertical axis
to match our results in the vicinity of 0.1 $M_\odot$.}
\label{fig8}
\end{figure}

The estimated mass function itself is plotted as a histogram in Figure 
\ref{fig8}.  Sources of error in this plot include errors
in mass estimation and misclassification as discussed in Section 5.
The results given in the Appendix suggest that
mass values have been estimated to an accuracy of a factor of $\sim2-3$.
The dashed line in the plot represents an estimate of the mass function
based on the assumption that all inferred cluster members are genuine.
We have corrected this result for likely contamination by
foreground and background objects by making use of the results
from the ``exterior" region, assuming that all inferred cluster
members in that region are spurious.  Specifically, we
constructed a separate mass function for
the ``exterior" region, scaled it by the cloud:exterior background source
count ratio (1.32) estimated in Section 5, and subtracted the scaled 
histogram from the mass function of the ``cloud" region.  The result is
indicated by the solid line in the figure.  

The mass function, within the limits of uncertainty indicated in Figure
\ref{fig8}, is consistent with the relatively flat distribution found 
previously in $\rho$ Oph for masses in excess of a few hundredths of a 
solar mass \citep{com93}, but departs significantly from this behavior for
lower masses.  Specifically,
our results suggest an order of magnitude increase in the number of objects
at $\sim0.003\,M_\odot$ with respect to that at $\sim0.1\,M_\odot$.
The actual increase may be even larger than this, however;
since some parts of the cloud have $A_V$ in excess of 100 magnitudes
(see Figure \ref{fig7}), we may be missing some low-mass objects
more deeply embedded in the cloud.  The apparent flattening-out of the
distribution below 0.002 $M_\odot$ is probably due to the sensitivity cutoff 
of the observations, as suggested by Figure \ref{fig7}. However, we cannot
rule out the possibility of an actual cutoff in the mass distribution.

Our inferred mass function is consistent with that found 
for $\sigma$ Orionis \citep{cab07,bih09} based on broad-band SED information 
using techniques somewhat similar to those used here.  They similarly 
found an order-of-magnitude increase in the mass function 
between 0.1 and 0.006 $M_\odot$; the \citet{cab07} results have been overplotted
on our Figure \ref{fig8} for comparison. In addition \citep{bih09}
found a possible hint of a turnover at $\sim0.004$ $M_\odot$,
also reminiscent of Figure \ref{fig8}.

\section{Discussion}

We have cross checked our results against previous work by searching
published lists of spectroscopically confirmed brown dwarfs
in the $\rho$ Oph cloud core region, but could only find one such case
within our field.  This is GY 204, listed by \citet{nat02} as 
an M6 brown dwarf with a temperature of 2700 K and a mass of 40--80 $M_J$.  
It coincides with object \#60 in Table 1
for which we had estimated $T_{\rm eff}=2888\pm340$ K and a mass of 
$59^{+89}_{-35}\,M_J$,
in agreement with the spectroscopic observation.
We were able to match the majority of the brighter cluster member candidates
to previously-known sources;  for the 57 cluster-member candidates
brighter than $K_s=14$, all but one corresponded to objects listed in the
SIMBAD database. Of those 56, all were consistent with being young objects 
associated with the $\rho$ Oph cloud.  This provides some additional confidence
in the assigned cluster membership.

Our results suggest that the mass function for the $\rho$ Oph cloud core
is similar to that of another young cluster,
$\sigma$ Orionis \citep{bih09}.  A key feature is
the increasing abundance with decreasing mass;
this may be a
consequence of the fragmentation of larger star-forming cores.  

The distributions of $J\!-\!H$ and $H\!-\!K_s$ colors for the lowest-mass 
objects ($\stackrel{<}{_\sim} 0.005\,M_\odot$) show a progressive blueward skew 
with decreasing flux, which we interpret in terms of decreasing dust extinction
with decreasing mass.  That is not to say that all low-mass objects have low
extinction, but rather that {\em some\/} of them make it out to the front edge
of the cloud where we see them unextincted. Such behavior might be
expected based on models in which the lowest-mass members of the cluster
have been dynamically ejected from a formation site deep in the cloud.
It is consistent with the timescales involved;
dynamical simulations suggest brown dwarf 
ejection velocities of the order of a few km s$^{-1}$ \citep{reip01},
adequate to traverse the $\rho$ Oph main cloud, of size $\sim1$--2 pc
\citep{wil05}, within the assumed 1 Myr age of the cloud.  

The spatial distribution of our low-mass ($M<5M_J$) brown dwarf candidates
is distinctly different from that of the higher-mass
T Tauri stars and YSOs, as shown in Figure \ref{fig5}.  One of the
key differences is that, surprisingly, the low-mass candidates are {\em less\/}
dispersed than the T Tauri stars, contrary to the expectations of the
ejection model.  Another key difference is that most
of the low-mass candidates are concentrated in the northernmost of the 
two dust filaments in this field ($\delta\sim-24.5^\circ$) and
are absent from the southern filement ($\delta\sim-24.7^\circ$), even
though the latter contains a somewhat higher concentration of T Tauri stars.
This contrasts with the findings of \citet{luh06},
whose study of the Taurus star-forming region indicated no significant 
difference between the distributions of stars and brown dwarfs.  
The situation is, however, somewhat reminiscent of spatial segregation 
effects in Taurus found by
\citet{guieu07}, whereby brown dwarfs with disks are preferentially located 
in one particular filament; no such segregation was evident
for the T Tauri stars.  Since the presence of a disk suggests
youth, the spatial segregation might be interpreted in terms of age 
differences between different aggregates of objects in the region.  
Similarly, in the case of $\rho$ Oph, the relatively compact 
aggregate of low-mass candidates in the northern filament in Figure \ref{fig5}
may have resulted from a star formation event more recent than for some or
all of the T Tauri stars.  Therefore, the spatial compactness of the 
low-mass aggregate does not necessarily argue against the ejection model.

\citet{alv10} have recently compared our photometry with their own
data, obtained in 2006, for the subset of 7 sources observed 
spectroscopically by \cite{mar10}, and found discrepancies in 3 cases.
In particular, they found \#4450 to be 1.42 magnitudes fainter than 
our estimate of $K_s=17.71$.  After further examination of our images,
we find the source to be slightly extended, with 
FWHM $\sim2''-3''$.  Its estimated flux will therefore depend to some
extent on the aperture or beamsize used. We have estimated its $K_s$-band 
aperture magnitude from the 2MASS Deep Field data using apertures of
various radii, making appropriate corrections for truncation
of the point-source response, and obtain $K_s = 18.34\pm0.15$ in a 
$1.5''$ aperture. We have compared this result with that obtained from
the $K$-band ``peak-up" image during the spectroscopic observations of 
\citet{mar10}, which yield $K=18.57\pm0.15$ for a $1.5''$ aperture; the image 
was unfortunately too noisy for larger apertures.  The two magnitudes
are nevertheless consistent within the error bars, and therefore suggest 
the absence of any significant source variability over the intervening
$\sim10$-year time span.  There remains a significant discrepancy with
the value $K_s=19.14\pm0.2$ estimated by 
\citet{alv10}, but we suspect that the difference can be attributed to
the smaller effective beamsize ($0.4''-0.8''$) in the latter observations, 
which could lead to flux underestimation for an extended source. 
Object \#4450 is not unique in this regard---we have found a number of
other cases in which our sources are slightly extended, and hypothesize that
we may be seeing the effects of scattering from remnant infalling dust 
envelopes surrounding the brown dwarf candidates.
It is not clear what, if any, effect such cases would have on our
temperature estimates, but we note that our SED fit for \#4450 yielded
an effective temperature in complete agreement with the spectroscopic value
from \citet{mar10}.

Our conclusions are based on fits to broad-band SEDs which are
subject to the uncertainties that we have discussed. Verification
must await spectroscopic observations in order to confirm the nature of 
individual objects and to better constrain their parameters.  
Nevertheless, SED fitting can play an important role in gathering statistics 
over wider areas of the sky, which is important to do because the mass 
function is known to vary from region to region---between 
different star-forming clouds \citep{evans91} and even within the same cloud 
\citep{bar97}.  Such studies will be aided by 
upcoming surveys, particularly the Wide-Field Infrared Survey Explorer (WISE) 
in conjunction with shorter-wavelength data from the UKIRT Infrared Deep Sky 
Survey (UKIDSS) and the Visible and Infrared Survey Telescope for Astronomy
(VISTA).  Additional complementary data, consisting of optical and far-red 
photometry, will soon
be available from the Pan-STARRS-1 and Sky-Mapper survey telescopes, and
will cover a larger area of sky than UKIDSS and VISTA.

\acknowledgments

We thank the referee for helpful comments and suggestions.
We also thank Tim Thompson for making the IRAC mosaic images, and John Stauffer 
and Luisa Rebull for helpful discussions.  
The research utilized data products from 2MASS, a joint project of the 
University of Massachussetts and IPAC/Caltech, and also archival data from 
{\em Spitzer\/}, operated by JPL/Caltech under contract to NASA. Use was also 
made of the SIMBAD database, operated at CDS, Strasbourg, France.
The work was carried out at IPAC/Caltech and was supported by a grant from 
the NASA Astrophysics Data Analysis Program. 

\section*{APPENDIX: Validation of model-fitting technique based on
spectroscopically-confirmed brown dwarfs}

We have evaluated the performance of our model-fitting procedure using
photometric data for cases of spectroscopically-confirmed brown dwarfs and
young stellar objects taken from the literature. Our selection criteria were:
\begin{itemize}
 \item[1.] The object must have been spectroscopically confirmed.
 \item[2.] The published photometry must include at least four of
 $J,\,H,\, K_s$, [3.6] (or $L$), and [4.5] (or $M$).
\end{itemize}

\begin{figure}
\epsscale{0.4}
\plotone{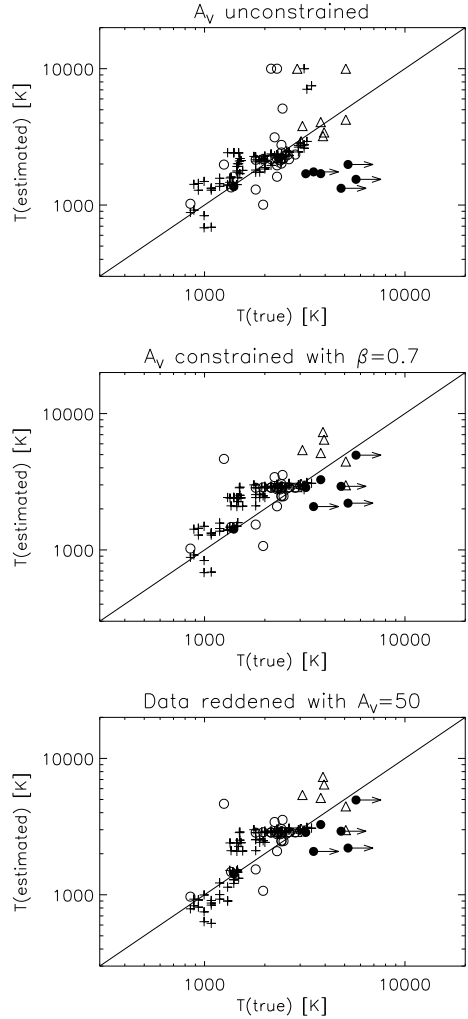}
\caption{Evaluation of our technique for temperature estimation using
photometric data for cases of spectroscopically confirmed brown
dwarfs and young stellar objects taken from the literature.  
The plots show the effective temperature estimated from our SED fits
as a function of the spectroscopically-determined value, as follows: 
{\em Top:\/}
no constraint on $A_V$; {\em Middle:\/} $A_V$ constrained with $\beta=0.7$
(see text); {\em Bottom:\/} the effect of artifically applying an additional
50 mag of visual extinction. Open circles represent young (1-10 Myr) brown dwarfs using
data from \cite{cab07}, \citet{luh05a,luh05b,luh07}, 
\citet{mart01}, \citet{riaz06} and \citet{zap00}; filled circles represent
the 7 young objects in $\rho$ Oph observed by \citet{mar10}; 
crosses represent field brown dwarfs using data compiled by \citet{pat06};
open triangles represent T Tau stars from \citet{wil05} and \citet{gat06}. }
\label{fig9}
\end{figure}

\begin{figure}
\epsscale{1.0}
\plotone{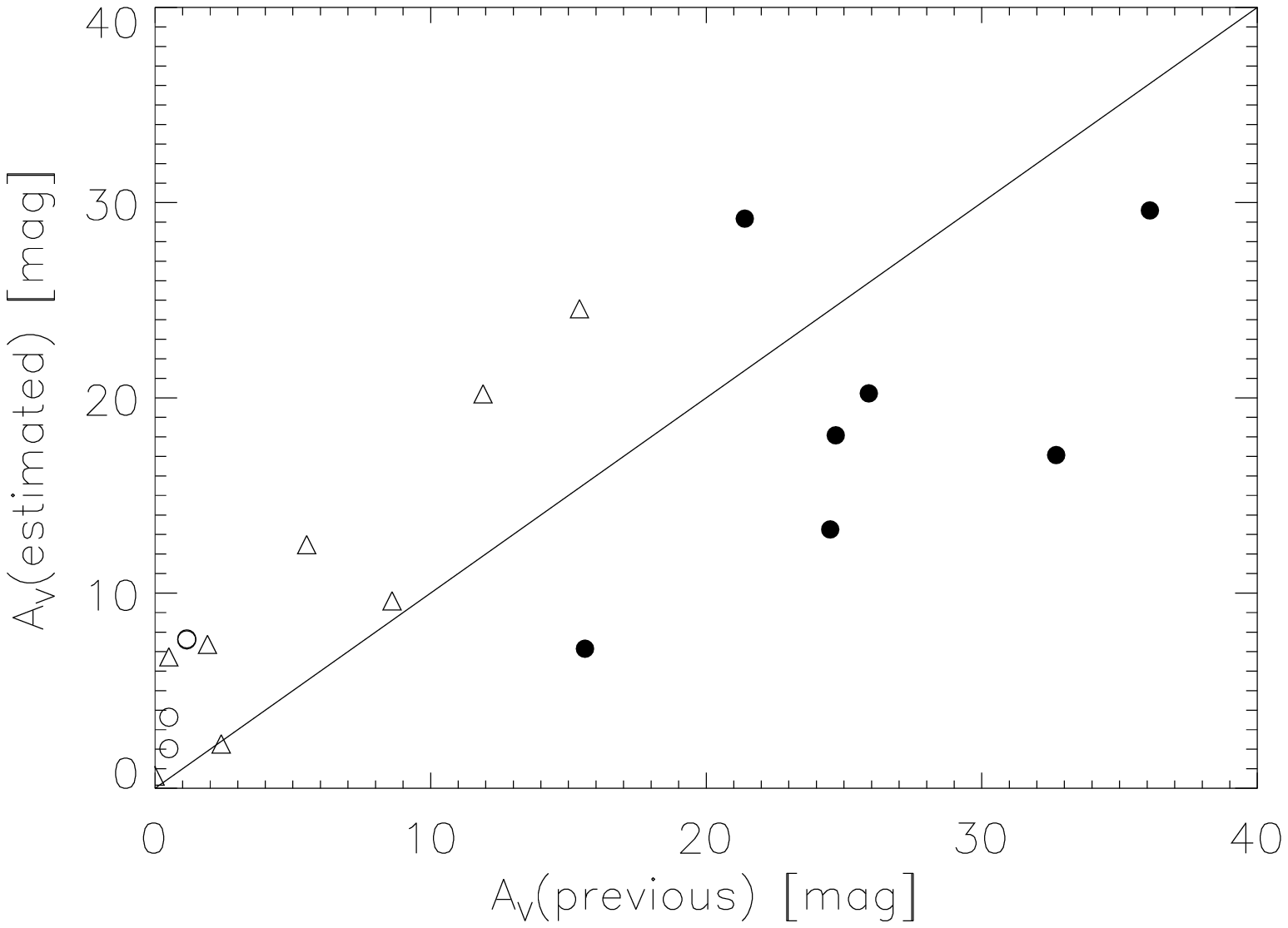}
\caption{Comparison of our estimated $A_V$ with previously 
published values. The symbol convention is the same as for Figure 9.}
\label{fig10}
\end{figure}

The selected data then consisted of 4- or 5-band photometric
measurements of 84 field brown dwarfs (ages 0.2--10 Gyr),
25 young brown dwarfs (ages $\sim1$--10 Myr), the 7 young objects in
$\rho$ Oph observed by \citet{mar10}, and all 8 of the 56 cross-identified 
SIMBAD objects for which a spectral type was available.  Our model fitting 
procedure was run on these data in a manner identical to that used in 
Section 4, i.e., a maximum likelihood fit to three unknowns
($T_{\rm eff}$, $\alpha$, $A_V$) using the same assumed age (1 Myr) for the
COND and DUSTY models. In each case, the SED-estimated temperatures were
compared with the published spectroscopic values.  In cases for which the
latter had not been specified, it was inferred from the published
spectral type using Figure 7 of \citet{kir05}. 
The results are presented in Figure \ref{fig9}, the
top panel of which shows the results obtained without the use
of the $A_V$ constraint discussed in Section 4, i.e., $\beta=0$.
The RMS difference between the estimated and spectroscopic values of
$\log T_{\rm eff}$ for that case was 0.18, corresponding to a percentage
error of 51\% in the estimated value of $T_{\rm eff}$.

We have optimized $\beta$ by minimizing the RMS difference in 
$\log T_{\rm eff}$ as a function of $\beta$.  In varying $\beta$ from 0 to
1.5, we find that the RMS difference goes through a well-defined minimum
of 41\%
at $\beta=0.7$, and back up to 51\% again; accordingly we select $\beta=0.7$
as the optimum value.  The middle panel in Figure \ref{fig9} shows
the result. It is apparent that without the $A_V$ constraint,
6 of the 7 \citet{mar10} objects
would have been incorrectly assigned temperatures below 2000 K.
A corresponding plot of our $A_V$ estimates versus the previously
published values, where available, is shown in Figure \ref{fig10}.

In order to assess the sensitivity of our temperature estimation technique
to the degree of extinction, we have artificially reddened the data by
various amounts and repeated the estimation procedure.  We find that
the applied extinction produces very little perturbation in estimated
temperature.  As an example, the bottom panel of Figure \ref{fig9}
shows the effect of an applied $A_V$ of 50 mag.

For each estimated temperature, the models provide
a corresponding mass which can be compared with the 
spectroscopically-estimated mass. For the young brown dwarfs, the 
RMS difference in
log mass ($\log_{10}M$) between our estimates and previous spectroscopic 
determinations was 0.41, corresponding to an average error of a factor
of $\sim2-3$ in mass estimation.  

Comparison between the uncertainties in $T_{\rm eff}$, $A_V$ in 
Table 1 and the scatter in these corresponding quantities in Figure
\ref{fig9} indicates that the true uncertainties are much larger than
the formal errors of maximum likelihood estimation.  The reason is
that the latter represents the effect of random noise and does not take
into account various sources of model error, which includes uncertainties
in age, the reddening law, and the photospheric models themselves.
The RMS residuals from the above evaluation therefore provide a more
realistic assessment of the true uncertainties in the quantities estimated
from the SED fits.

\clearpage

 \begin{deluxetable}{ccccccccccc}
\tabletypesize{\scriptsize}
\rotate
\tablecaption{Photometry and estimated effective temperatures, visual absorption,
and masses of candidate cluster members in the $\rho$ Oph cloud core.}
\tablewidth{0pt}
\tablehead{
\colhead{Obj\#} & \colhead{RA} & \colhead{Dec} & \colhead{$J$} & 
\colhead{$H$} & \colhead{$K_s$} & \colhead{[3.6]} &  \colhead{[4.5]} & \colhead{$T_{\rm eff}$} & 
\colhead{$A_V$} & \colhead{$M/M_\odot$} 
}
\startdata
  10 &  16 27 19.51 &  -24 41 40.2 &    9.401 (0.006) &    8.654 (0.006) &    8.404 (0.006) &    8.569 (0.030) &    8.184 (0.022) &   5135 (1071) &  2.3 (1.6) & 1.849  \\
  14 &  16 27 15.13 &  -24 51 38.7 &   10.660 (0.006) &    9.813 (0.006) &    9.465 (0.006) &    8.949 (0.027) &    8.354 (0.022) &   3033 (296)  &  3.5 (1.3) & 0.112  \\
  17 &  16 27  9.10 &  -24 34  8.0 &   12.662 (0.006) &   10.264 (0.006) &    8.927 (0.006) &    7.781 (0.046) &    6.876 (0.026) &   7319 (2872) &  24.6 (1.4)& 2.530  \\
  20 &  16 27 13.74 &  -24 18 16.7 &   12.296 (0.006) &   10.258 (0.006) &    9.241 (0.006) &    8.548 (0.029) &    7.632 (0.024) &   5874 (1499) &  18.1 (1.5)& 1.994  \\
  21 &  16 27 32.86 &  -24 53 45.4 &   10.066 (0.006) &    8.793 (0.006) &    8.342 (0.006) &    8.425 (0.036) &    8.187 (0.023) &   4426 (430)  &  4.7 (1.1) & 0.687  \\
  22 &  16 27 17.08 &  -24 47 11.0 &   11.847 (0.006) &   10.173 (0.006) &    9.496 (0.006) &    9.156 (0.026) &    9.026 (0.023) &   4451 (475)  &  9.6 (1.2) & 0.711  \\
  23 &  16 27  5.17 &  -24 20  7.6 &   12.716 (0.006) &   10.448 (0.006) &    9.351 (0.006) &    8.877 (0.029) &    8.337 (0.023) &   4382 (404)  &  16.0 (1.1)& 0.646  \\
  24 &  16 27 27.38 &  -24 31 16.5 &   12.364 (0.006) &   10.378 (0.006) &    9.321 (0.006) &    8.653 (0.030) &    8.015 (0.023) &   5348 (1258) &  17.1 (1.5)& 1.983  \\
  25 &  16 26 56.77 &  -24 13 51.4 &   12.390 (0.006) &   10.388 (0.006) &    9.323 (0.006) &    8.250 (0.041) &    7.582 (0.031) &   8008 (8032) &  20.6 (2.7)& 2.814  \\
  27 &  16 27 30.85 &  -24 47 26.7 &   12.186 (0.006) &   10.395 (0.006) &    9.485 (0.006) &    8.972 (0.027) &    8.661 (0.023) &   4643 (803)  &  13.1 (1.6)& 0.971  \\
  28 &  16 27 10.32 &  -24 19 18.7 &   11.988 (0.055) &   10.456 (0.025) &    9.759 (0.021) &    9.245 (0.065) &    9.007 (0.053) &   5127 (1180) &  11.2 (1.7)& 1.842  \\
  30 &  16 27 22.93 &  -24 17 57.3 &   13.314 (0.006) &   10.695 (0.006) &    9.397 (0.006) &    8.771 (0.027) &    8.187 (0.022) &   4352 (346)  &  19.5 (1.0)& 0.620  \\
  32 &  16 27  5.91 &  -24 59 37.8 &   11.526 (0.006) &   10.436 (0.006) &   10.068 (0.006) &        \nodata   &    9.877 (0.022) &   4804 (868)  &  5.2 (1.8) & 1.297  \\
  34 &  16 27  4.52 &  -24 42 59.5 &   12.016 (0.006) &   10.491 (0.006) &    9.829 (0.006) &    9.389 (0.028) &    9.176 (0.022) &   4812 (897)  &  10.1 (1.6)& 1.314  \\
  35 &  16 27  1.63 &  -24 21 36.9 &   14.303 (0.006) &   11.057 (0.006) &    9.397 (0.006) &    8.546 (0.027) &    8.116 (0.023) &   4256 (254)  &  27.2 (1.3)& 0.519  \\
\enddata
\tablecomments{ The columns represent the object number, RA and Dec position 
(J2000), the magnitudes at the five bands indicated, the 
estimated effective temperature ($T_{\rm eff}$ [K]), visual absorption ($A_V$), and mass ($M$).  
Uncertainties are indicated in parentheses, with the exception of RA and Dec
($0.5''$ each) and $M$ (a factor of $\sim2-3$). 
Table \ref{tbl-1} is published in its entirety (165 objects) in the 
electronic edition of the {\it Astrophysical Journal}.  A portion is 
shown here for guidance regarding its form and content.} 
\label{tbl-1}
\end{deluxetable}

\end{document}